\documentclass[prd,showpacs,twocolumn,showkeys,floatfix,%
superscriptaddress]{revtex4}
\usepackage{graphicx}
\begin{document}
\title{Multi--scalar black holes with contingent primary hair:
Mechanics and stability}
\author{Salvatore Mignemi}
\email{smignemi@unica.it}
\affiliation{Dipartimento di Matematica, Universit\`a di Cagliari,
viale Merello 92, 09123 Cagliari, Italy}
\affiliation{INFN, Sezione di Cagliari, Cittadella Universitaria,
09042 Monserrato, Italy}
\author{David L. Wiltshire}
\email{d.wiltshire@phys.canterbury.ac.nz}
\affiliation{Department of Physics and Mathematical Physics,
University of Adelaide, Adelaide, S.A. 5005, Australia.}
\affiliation{Department of Physics and Astronomy,
University of Canterbury, Private Bag 4800, Christchurch, New Zealand}
\altaffiliation[Present address]{}

\begin{abstract}
We generalize a class of magnetically charged black holes holes
non-minimally coupled to two scalar fields previously found by one of us
to the case of multiple scalar fields.
The black holes possess a novel type of primary scalar hair, which we
call a {\it contingent primary hair}: although the solutions possess
degrees of freedom which are not completely determined by the other
charges of the theory, the charges necessarily vanish in the absence of
the magnetic monopole. Only one constraint relates the black
hole mass to the magnetic charge and scalar charges of the theory.
We obtain a Smarr-type thermodynamic relation, and the first law of black
hole thermodynamics for the system.
We further explicitly show in the two--scalar--field case that,
contrary to the case of many other hairy black holes, the
black hole solutions are stable to radial perturbations.
\end{abstract}
\keywords{Black holes, scalar fields, hair, stability;\qquad
[{\bf Report no}: ADP-00-46/M94, hep-th/0408215]}
\maketitle


\def\bd{\begin{displaymath}}\def\ed{\end{displaymath}}
\def\beq{\begin{equation}}\def\eeq{\end{equation}}
\def\bea{\begin{eqnarray}}\def\eea{\end{eqnarray}}
\def\nn{\nonumber}\def\lb{\label}

\def\de{\partial}\def\e{{\rm e}}\def\ka{\kappa}
\def\const{{\rm const}}\def\ha{{1\over2}}
\def\efs{(\e^{-2\Phi}+\e^{-2\Psi})}
\def\ef{\e^{-2\Phi}}\def\es{\e^{-2\Psi}}
\def\dG{\delta\Gamma}\def\dL{\delta\Lambda}
\def\dP{\delta\Phi}\def\dS{\delta\Psi}
\def\rp{r_+}\def\rq{r_-}
\def\conn{\left({\Gamma'-\Lambda'\over2}+{2\over R}\right)}
\def\cont{\left({\Gamma'-\Lambda'\over2}+{3\over R}\right)}
\def\cons{{\dot\Lambda-\dot\Gamma\over 2}}
\def\elg{\e^{\Lambda-\Gamma}}\def\egl{\e^{\Gamma-\Lambda}}
\def\elf{\e^{\Lambda-2\Phi}}\def\els{\e^{\Lambda-2\Psi}}
\def\bu{{\bf u}}\def\bA{{\bf A}}
\def\goesas{\mathop{\sim}\limits} \def\etal{{\it et al.}}
\def\LA{\Lambda}\def\OM{\Omega}\def\al{\alpha}\def\be{\beta}
\def\Si{\Sigma}\def\rh{\rho}\def\th{\theta}\def\ep{\epsilon}
\def\rarr{\rightarrow} \def\mfac{\sqrt{\frac23}\,m} \def\CC{{\cal C}}
\def\const{\hbox{const}} \def\OO{{\rm O}}
\def\AH{{\cal A}_{\lower2pt\hbox{$\scriptstyle\cal H$}}}
\def\chiH{\chi_{\lower2pt\hbox{$\scriptstyle\cal H$}}}
\def\RR{{\cal R}} \def\pt{\partial} \def\si{\sigma}
\def\scrscr{\scriptscriptstyle} \def\ph#1{\phi^{#1}}
\def\la#1{{\lambda_{#1}}} 
\def\gg#1{{g_#1}} 
\def\SUM#1{\sum_{a=#1}^N} \def\sc#1{{\Sigma_{#1}}}
\def\eph{\e^{-2\gg{a}\ph{a}}} \def\Qeph{{Q_a}^2\eph} \def\Ra#1{R_{\scrscr#1}}
\def\dd{{\rm d}} \def\uu{\e^{2U}} \def\rmm{(r-r_-)} \def\rmp{(r-r_+)}
\def\Ddr{{\dd\phantom{r}\over\dd r}} \def\phm#1{\ph{#1}_{\scrscr\infty}}
\def\BQ{{\bar Q}}\def\bQ{\BQ^2} \def\X#1{_{\lower2pt\hbox{$\scrscr#1$}}}
\def\De{\Delta}
\section{Introduction}

In classical general relativity, no hair theorems \cite{Be} impose strong
constraints on the possibility of obtaining black hole solutions of
the Einstein equations coupled to non-trivial scalar fields.
A crucial ingredient for their proof is that the scalars be minimally
coupled to gravity and other fields. When this condition is relaxed new
possibilities emerge for evading the no hair theorems.

One of the first investigations was undertaken by Bekenstein \cite{Be1},
who attempted to show that a non-trivial black hole solution exists
for a conformally coupled scalar field. Although the scalar field diverges
at the horizon, Bekenstein argued that this solution nevertheless admits a
physical interpretation. Unfortunately, a more detailed analysis of the
problem casts serious doubt on this \cite{SZ}. We will therefore take the
view that scalar fields should be regular at the horizon for genuine black
hole solutions.

Non-trivial scalar hair is possible, however, when black holes are coupled
to scalar fields with non-linear self-interactions. Such solutions were
first found in the case of gravity coupled to the Skyrme model \cite{LM},
and subsequently for the Einstein-Yang-Mills-Higgs model \cite{EYMH},
and other generalizations. The large literature on this topic has been
recently reviewed in ref.\ \cite{VG}. What appears to be characteristic of
the class of black holes with non-linear self-interactions is that
the scalar fields fall off very rapidly at spatial infinity -- e.g.,
exponentially fast -- and hence the asymptotic scalar charges vanish.
The scalar hair is characterized instead by non-trivial charges on the
horizon.

A third possibility is that of minimally coupled scalar fields with
potentials, $V(\phi)$, which violate the dominant energy condition (DEC)
\cite{BL,BS,NS}. In single scalar field models, the DEC is equivalent
to $V(\phi)\ge0$, and thus one must choose potentials for which $V(\phi)<0$
for some field values. Examples of asymptotically flat black hole
solutions have been found for cases in which $V(\phi)<0$ everywhere \cite{BS},
and for cases in which $V(\phi)$ possesses at least one global minimum at
negative values \cite{BL,NS}. Both analytic \cite{BL,BS} and numerical
\cite{BL,NS} examples are known.

A final possibility is to consider black holes with scalar fields which are
non-minimally coupled to gauge fields. Such models have been extensively
investigated because they arise naturally in Kaluza-Klein theories
and in the effective low-energy limit of string theory, where the dilaton
plays a non-trivial role. In all these models one can find black hole
solutions with non-zero scalar charges at spatial infinity \cite{DM,GM}.
An analogous phenomenon can be shown to take place also in the
pure dilaton-gravity sector of effective string theories,
when one takes into account the coupling of the dilaton to gravity
via Gauss-Bonnet terms \cite{MS}.

For this class of non-minimally coupled models with non-zero asymptotic
scalar charges, however, the scalar charges in question are {\it not}
independent parameters, but in all cases are a given function of the other
asymptotic charges which characterize the solution, namely the ADM mass and
the electric and magnetic charges etc. As a result such scalar charges have
been called {\it secondary hair} by the authors of Refs.\ \cite{CPW1,CPW2},
to distinguish them from the theoretical possibility of a {\it primary
hair}, namely an asymptotic scalar charge which is completely independent of
the other charges.

In light of all the known solutions \cite{VG} one may therefore conjecture
\cite{CPW2} that a weaker form of the no-hair property still holds, namely
that for theories which satisfy the DEC black hole solutions can be classified
by a small number of parameters, which in the case of asymptotic charges
include only conserved charges such as mass, angular momentum and gauge
charges, but no asymptotic scalar charges.

Even this no hair conjecture does not appear to be valid in
general, however. In a recent paper by one of us \cite{SM1} the
properties of magnetically charged black holes coupled to a
dilaton, $\Phi$, and an additional modulus field, $\Psi$,
according the the field equations generated by the 4-dimensional
action \cite{notnote} \bea
S={1\over4}\int\dd^4x\sqrt{-g}\Biggl\{&-&\RR+2\pt^\mu\Phi\pt_\mu\Phi
+2\pt^\mu\Psi\pt_\mu\Psi\nonumber\\
&-&\left[\e^{-2\Phi}+(\la2)^2\e^{-2q\Psi}\right]
F_{\mu\nu}F^{\mu\nu}\Biggr\},\nonumber\\ \lb{action2}\eea were
studied using techniques from the general theory of dynamical
systems, which have been previously applied to static spherically
symmetric solutions of gravity coupled to scalar fields in a
number of contexts \cite{MW}. In Ref.\ \cite{SM1} it was shown
that the regular black hole solutions were parameterized by an
additional degree of freedom in addition to the mass, $M$, and
magnetic charge, $Q$, and it was conjectured that this degree of
freedom could be considered to be a ``primary scalar hair''.

An important issue in this context is that of stability of the
hairy solutions. The physical relevance of the solutions would in
fact be spoiled by the presence of instabilities. Instabilities are known
to be present in the case of Einstein--Yang--Mills models \cite{stab,VG2}
and the DEC--violating solutions for which the stability issue has
been most thoroughly studied \cite{NS}. Dilaton black holes with
a single scalar do not appear to suffer from instability problems
\cite{HW,KW}; however, they possess only secondary hair.

It is the aim of the present paper to clarify and extend the results of
Ref.\ \cite{SM1} by determining various relations satisfied by the charges,
and discussing the stability of the solutions. We will find
in particular, that although the scalar degree of freedom cannot be considered
to be a primary hair in the strictest sense, since it must necessarily vanish
if the electromagnetic field vanishes, the regular static spherically
symmetric black hole solutions are not completely specified by their mass and
magnetic charge. The solutions therefore are potentially of considerable
physical interest as the only known static spherically symmetric solutions
with non-trivial scalar charges at spatial infinity which are not
completely determined by the other asymptotic charges of the theory, and
therefore provide a counter-example to some forms of the no-hair conjecture.
We will show that the solutions are stable, adding significance to
their interpretation.

Intuitively, one might attempt to understand the physical basis of the
no-hair conjecture as arising from the fact there is no scalar equivalent
of Gauss's law to give a conserved charge. The presence of the
electromagnetic field supports the scalar charge in the case of models with
a secondary scalar hair since the scalar coupling enters into the equivalent
of Gauss's law. In the solutions described here the scalar charges are also
supported by the presence of the electromagnetic field. However, they are
not entirely determined by it.

It is quite consistent with past usage to adopt the terminology
``primary hair'' for the additional degrees of freedom which arise
in the model \cite{SM1}. However, the fact that the scalar charges are not
entirely independent of the electromagnetic field suggests a new terminology
might be appropriate, as a way of capturing the finer distinctions that
the present model has revealed. We will therefore refer to the scalar
charges which could theoretically exist in the
absence of additional non-zero gauge charges as
{\it elementary primary hair}, as would be consistent with
the original aims of the first ``no hair'' theorems \cite{Be}.
The additional degrees of freedom which arise in the present model
could by contrast be deemed to be a {\it contingent primary hair}.

\section{Multi--scalar black hole solutions}

Rather than restricting our attention to solutions of the field equations
obtained from varying the action (\ref{action2}), we will instead consider
the somewhat more general action for $N$ scalar fields coupled to a single
$U(1)$ Abelian gauge field according to
\bea S={1\over4}\int\dd^4x\sqrt{-g}\Biggl\{&-&\RR+2\SUM1\pt^\mu\ph{a}\pt_\mu
\ph{a}\nonumber\\ &-&\SUM1(\la{a})^2\eph
F_{\mu\nu}F^{\mu\nu}\Biggr\},\nonumber\\ \lb{action}\eea
where $\gg{a}\ne0$, since the problem is not significantly more complicated.
Rather than using the coordinates which were exploited in \cite{SM1}, we
will take static spherically symmetric solutions with coordinates
\beq \dd s^2=-\e^{2U}\dd t^2+\e^{-2U}\dd r^2+R^2\dd{\OM_2}^2,
\lb{coorda} \eeq
where $U=U(r)$ and $R=R(r)$, and $\dd{\OM_2}^2$ is the standard round metric
on a $2$-sphere. We take $R>0$ without loss of generality. We are primarily
interested in black holes solutions with at least one regular horizon which
are asymptotically flat as $R\rarr\infty$.

If we assume that the gauge field is given by a magnetic monopole
configuration with components
\beq F_{\hat\th_1\hat\th_2}={Q\over R^2}\,\ep_{\hat\th_1\hat\th_2}
\lb{mag}\eeq
in an orthonormal frame then the Maxwell--type equation,
\beq \SUM1(\la{a})^2\pt_\mu\left[\eph\sqrt{-g}\,F^{\mu\nu}\right]=0,
\lb{Max}\eeq
is satisfied identically,
and the remaining field equations obtained from variation of the
action (\ref{action}) take the form
\bea
\left[R^2\uu{\ph{a}}'\right]'&=&-{\gg{a}\over R^2}\Qeph\,,\lb{feaA}\\
R^2\left(\uu\right)''+2RR'\left(\uu\right)'&=&{2\over R^2}\SUM1\Qeph,
\lb{feaB}\\ {R''\over R}&=&-\SUM1{\ph{a}}'^2,\lb{feaC}\\
\left[\uu\left(R^2\right)'\right]'=2&-&{2\over R^2}\SUM1\Qeph\,
\lb{feaD}
\eea
where
\beq Q_a\equiv\la{a}Q \eeq
and a prime denotes $\dd/\dd r$. By virtue of the Bianchi identity one of
Eqs.\ (\ref{feaA})--(\ref{feaD}) can be derived from the others.

One should note that in contrast to
many simpler models \cite{DM,GM,BMG} there is no simple duality relation
between magnetic and electric solutions in the theory, on account of the
multi--scalar exponential coupling term which multiplies the electromagnetic
part of the action (\ref{action}). Thus we cannot simply read off the
properties of solutions with an electric field in place of the monopole
ansatz (\ref{mag}). It might be tempting to think of the action (\ref{action})
as a special case of an alternative model in which each scalar field is
associated with an independent $U(1)$ gauge field. However, (\ref{action})
is not a simple truncation of such a model, since such an increase of the
number of $U(1)$ fields would lead to $N$ independent Maxwell--type equations,
each with its own single scalar coupling, rather than a single equation of
the form (\ref{Max}). Thus the present model can be expected to display
differences in comparison to models in which moduli and several independent
$U(1)$ fields are present, allowing for the choice of duality--preserving
combinations \cite{BMG,CC}.

\begin{figure}[htb]
\vbox{ \vskip 10 pt
\centerline{\scalebox{0.75}{\includegraphics{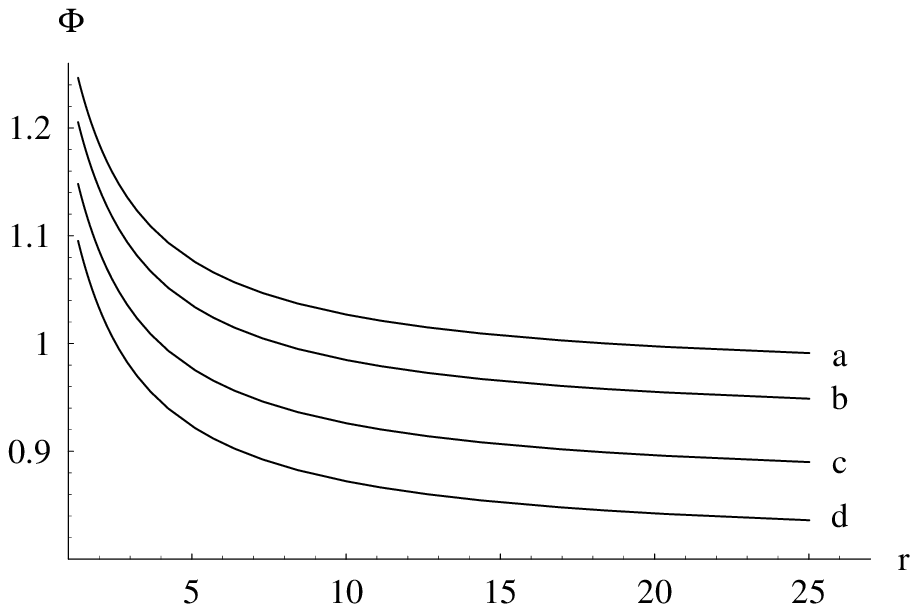}}} \vskip
10 pt \centerline{\scalebox{0.75}{\includegraphics{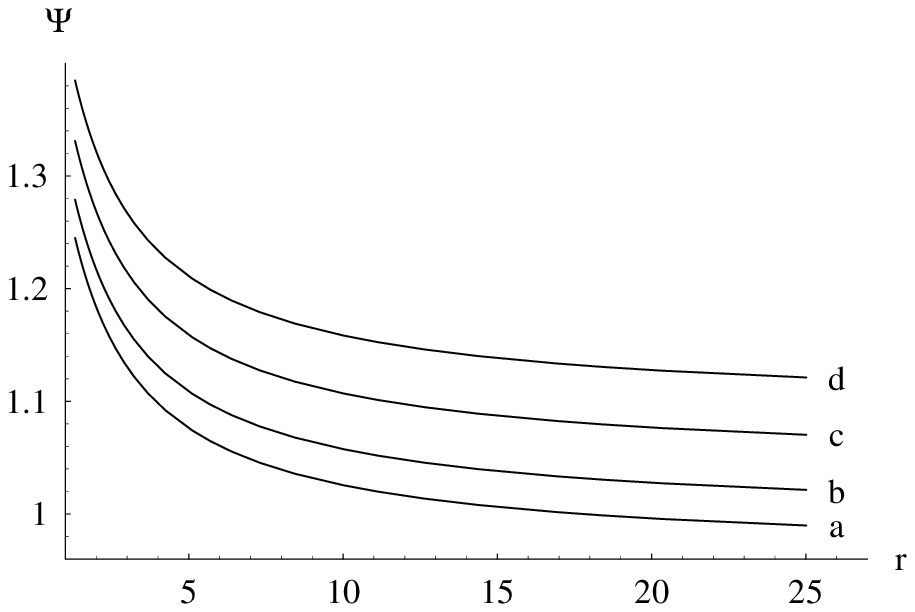}}}
\caption{%
{\sl The scalars $\Phi$ and $\Psi$, solutions of the model (1) with
$\lambda_2=1$, $q=1$, plotted for fixed values of $r_+$ and
$r_-$, and variable third independent parameter. Case (a)
corresponds to $\Si_\Phi=\Si_\Psi$, and the other cases are
ordered according to decreasing ratio $\Si_\Phi/\Si_\Psi$.}}}
\end{figure}

\subsection{Nature of horizons}

A straightforward proof by contradiction of the type used in Refs.\
\cite{PTW} may be used to establish that the solutions of
(\ref{feaA})--(\ref{feaD}) possess at most one regular horizon. We first
note that the sum of Eqs.\ (\ref{feaB}) and (\ref{feaD}) gives
\beq \left[\uu R^2\right]''=2 \lb{feaE}\eeq
which may be integrated to yield
\beq \uu R^2=r^2+\al r+\be,\lb{RU1}\eeq
for arbitrary constants $\al$ and $\be$. We assume that (\ref{RU1}) possesses
at least one real zero in order that there may exist at least one horizon
and thus rewrite (\ref{RU1}) as
\beq \uu R^2=\rmm\rmp,\lb{RU2}\eeq
where $\al=r_++r_-$ and $\be=r_-r_+$, and we may assume without loss of
generality that $r_-\le r_+$. We substitute (\ref{RU2}) into (\ref{feaA})
to obtain
\beq\left[\rmm\rmp{\ph{a}}'\right]'=-{\gg{a}\over R^2}\Qeph\,,
\lb{febA}\eeq

Let us now suppose that, $r_+\ne r_-$, both values $r=r_\pm$ correspond to
regular horizons, and that $\gg{a}>0$ for any one of the scalar fields.
Since the scalar is smooth at the horizon, if we evaluate (\ref{febA}) at
$r=r_+$ we obtain
\beq (r_+-r_-){\ph{a}}'(r_+)=-{\gg{a}\over R^2(r_+)}{Q_a}^2\e^{-2\gg{a}\ph{a}
(r_+)}\lb{phimin}\eeq
from which it follows that ${\ph{a}}'(r_+)<0$. Similarly, if we evaluate
(\ref{febA}) at $r=r_-$ we see that ${\ph{a}}'(r_-)>0$. Now given that
$\ph{a}(r)$ is assumed to be smooth, it follows that it must have a maximum
at an intermediate value $r=r_0$ such that $r_-<r_0<r_+$. However, if we
evaluate (\ref{febA}) at $r=r_0$ we obtain
\beq (r_0-r_-)(r_0-r_+){\ph{a}}''(r_0)=-{\gg{a}\over R^2(r_0)}{Q_a}^2
\e^{-2\gg{a}\ph{a}(r_0)}\eeq
from which it follows that ${\ph{a}}''(r_0)>0$, giving a minimum, which is a
contradiction. Thus if ${\ph{a}}$ is regular at $r=r_+$ it cannot also be
regular at $r=r_-$. The point $r=r_-$ should thus correspond to a curvature
singularity.

Similarly, if we assume that $\gg{a}<0$, then all the signs in the above
arguments are reversed but we still obtain contradiction. It therefore
follows that the solutions can at most possess one regular horizon, at
$r=r_+$.

It is also useful to note that if $\gg{a}>0$ then $\ph{a}$ must be
monotonically decreasing on the domain of outer communications
of regular black hole solutions, since if it reached a minimum at a finite
value $r_0>r_+$ then at such a point Eq.\ (\ref{phimin}) would once again
be true, but now with the implication that ${\ph{a}}''(r_0)<0$, again giving a
contradiction. Likewise if $\gg{a}<0$ then $\ph{a}$ is monotonically increasing
for $r\ge r_+$.

The function $R(r)$ must be monotonically increasing in the domain of outer
communications for either sign of $\gg a$, since by (\ref{feaC}) $R''<0$
at finite $r$ for any solutions with non-trivial scalars. This leaves a global
maximum as the only possible turning point for the function $R(r)$, but such
a choice would be inconsistent with asymptotic flatness, since by (\ref{RU2})
$R\goesas r$ as $r\rarr\infty$, given $\uu\rarr1$.

\subsection{Constraints on charges}

An additional first integral of the field equations may be extracted as
follows: take the difference of (\ref{feaB}) and (\ref{feaD}), eliminate
terms involving $\left(\uu\right)''$ using (\ref{RU2}), and eliminate terms
involving $\eph$ using (\ref{feaA}). Integrating the resulting equation one
obtains
\beq R^2\uu\SUM1{{\ph{a}}'\over\gg{a}}=RR'\uu-r+c\lb{Rp1}\eeq
where $c$ is an arbitrary constant. In order to obtain solutions which are
regular at the outer horizon, it is in fact necessary to choose $c=r_+$.
With this choice, and again using (\ref{RU2}), (\ref{Rp1}) may be
integrated to yield
\beq A_0\exp\left[\SUM1{\ph{a}\over\gg{a}}\right]={R\over r-r_-}\,,
\lb{phisum}\eeq
where $A_0$ is an arbitrary constant. From (\ref{phisum}) we see that $r=r_-$
will correspond to a singularity, as expected from above.

Unfortunately, it does not appear to be possible to obtain an analytic solution
in closed form to the remaining field equations. Since one of the $N$ scalar
equations (\ref{feaA}) can be eliminated with the use of (\ref{phisum}), and
since the function $\uu$ is given in terms of the function $R$ according to
(\ref{RU2}), we are left with one first order ODE for $R$, namely
\bea &&\hbox{\hskip-10pt}\left({R'^2\over R^2}+\SUM1{\ph{a}}'^2\right)(r-r_+)
(r-r_-)-2{R'\over R} \left(r-M\right)+1\nonumber\\ &&\qquad={1\over R^2}
\SUM1\Qeph\lb{febB}\eea
coupled nonlinearly to $N-1$ independent second order ODEs (\ref{febA}) for
the scalars. This is equivalent therefore to a system of $2N-1$
first order ODEs, which can be solved numerically.

Much useful analytic information about the solutions can be obtained in
relation to the values of the ADM mass, $M$, and the $N$ scalar charges,
$\sc{a}$, which correspond to the $\OO(r^{-1})$ terms in the asymptotic
expansions at spatial infinity,
\bea
\uu&=&1-{2M\over r}+{u_{\scrscr2}\over r^2}+\dots,\lb{asymA}\\
\ph{a}&=&\phm{a}+{\sc{a}\over r}+{\ph{a}_{\scrscr2}\over r^2}
+\dots,\lb{asymB}\\ R^2&=&r^2\left(1+{\Ra1\over r}+{\Ra2\over r^2}
+\dots\right)\;.\lb{asymC}\eea
The $\OO(r)$ coefficient, $\Ra1$, of the function $R^2$ is a gauge
parameter whose value fixes the choice of origin of the radial coordinate,
$r$. We will use this gauge freedom to set $$\Ra1=0.$$
Expanding Eq.\ (\ref{phisum}) at spatial infinity by use of (\ref{asymB})
and (\ref{asymC}) it follows from the leading order term that the
constant $A_0$ is related to the moduli vacuum charges, $\phm{a}$, according to
\beq A_0=\exp\left[-\SUM1{\phm{a}\over\gg{a}}\right]\,.
\lb{phimodsum}\eeq
Furthermore, if we also make use of (\ref{asymA}) it follows from the
next to leading order terms in (\ref{RU2}) and (\ref{phisum}) that the
following relations hold between the constants $r_\pm$ and the asymptotic
charges
\beq r_\pm=M\pm\left(M-\SUM1{\sc{a}\over\gg{a}}\right).\lb{rpm1}\eeq
The constraint that $r_+\ge r_-$ then yields the inequality
\beq \SUM1{\sc a\over\gg a}\le M,\lb{scineq}\eeq
which is saturated for the extreme solutions for which the horizon is
degenerate with the inner singularity.

With definitions of the asymptotic charges in hand we can now integrate
various field equations between the horizon, $r=r_+$, and spatial infinity
to obtain constraints on the charges. If the scalar equations (\ref{feaA})
are integrated on this interval, for example, we find that
\beq\sc{a}=\gg{a}{Q_a}^2\int_{r_+}^\infty\dd r{\eph\over R^2}\,.
\lb{screl}\eeq
We note that for solutions which are regular at the horizon, given that the
integrand of (\ref{screl}) is positive it follows that
\beq {\sc a\over\gg a}\ge0\lb{scsign}\eeq
for each scalar charge, and furthermore $\sc{a}=0$ if and only if $Q=0$. Thus
the charges $\sc{a}$ do not constitute an elementary primary
scalar hair according to the definition adopted in the Introduction.

At first sight one might be tempted to assume that Eqs.\ (\ref{screl})
provide constraints on all $N$ scalar charges, and that we are therefore
dealing with a system with purely secondary scalar hair. However, this
is {\it not} in fact the case since the lower limit of integration, $r_+$,
already depends on the ADM mass and scalar charges according to
(\ref{rpm1}).

The nature of the relations (\ref{screl}) is more readily understood
if we rewrite them in terms of functions $\tilde\ph{a}\equiv\ph{a}-\phm{a}$,
$a=1\dots N$, which have the leading order behaviour $\tilde\ph{a}(r)\goesas
\sc{a}/r+\dots$ at spatial infinity. We then see that equations
(\ref{screl}) can be rewritten as
\beq\phm{a}={1\over2\gg{a}}\ln\left[{\gg{a}{Q_a}^2\over\sc{a}}
\int_{r_+}^\infty\dd r{\e^{-2\gg{a}\tilde\ph{a}}\over R^2}\right]\,
\lb{screl2}\eeq
for $\sc{a}\ne0$. Since the bounds of integration are independent of
the moduli vacuum charges, we see that the relations (\ref{screl}) or
(\ref{screl2}) are constraints which determine the $\phm{a}$ in terms of
$Q$, $M$ and $\sc{a}$.

Let us turn to the question of whether there exist any constraints on
the scalar charges $\sc{a}$.
In fact, there appears to be only one additional constraint on the asymptotic
charges. This may be determined by noting that if one multiplies each of the
scalar equations (\ref{feaA}) by $2R^2\uu{\ph{a}}'$ and then takes the sum
of the resulting $N$ equations plus $\frac12R^2\left(\uu\right)'$ times the
difference of Eqs.\ (\ref{feaD}) and (\ref{feaE}), one obtains the
expression
\bea\Ddr&&\left\{\frac14\left[R^2\left(\uu\right)'\right]^2
+\SUM1\left[\left(R^2\uu{\ph{a}}'\right)^2\right]\right\}\nonumber\\
&&\qquad=\Ddr\left\{\uu\SUM1\Qeph\right\}\lb{diffconstraint}\eea
We may integrate Eq.\ (\ref{diffconstraint}) from $r=r_+$ to spatial
infinity, and use (\ref{RU2}) and (\ref{rpm1}) to obtain the following
constraint on the charges
\beq \SUM1\sc{a}^2+2M\SUM1{\sc{a}\over\gg{a}}-\left(\SUM1{\sc{a}\over\gg{a}}
\right)^2=\bQ,\lb{constraint}\eeq
where
\beq \bQ\equiv\SUM1\bQ_a\lb{bQ1}\eeq
with
\beq \bar Q_a\equiv\e^{-\gg{a}\phm{a}}Q_a=\la{a}\e^{-\gg{a}\phm{a}}Q\,.
\lb{bQ2}\eeq
The quantity $\bar Q$ can be thought of as
the magnetic monopole charge normalized by the weighted sum of the
moduli vacuum charges. On account of the possibility of different vacuum
moduli charges, the individual scalars can effectively ``see'' different
magnetic monopole charges, $\bar Q_a$.

Using Eq.\ (\ref{constraint}) we obtain the following expression equivalent to
(\ref{rpm1})
\beq r_\pm=M\pm\left[M^2+\SUM1\sc{a}^2-\bQ\right]^{1/2},\lb{rpm2}\eeq
and $\bQ$ is bounded above according to
\beq\bQ\le M^2+\SUM1\sc{a}^2.\lb{qineq}\eeq

The constraint (\ref{constraint}) reduces the number of independent scalar
charges to $N-1$. Do any further constraints remain to be found? In the
$N=2$ case of two scalar fields this cannot be the case, given the
numerical results of Fig.\ 1 and the results of the dynamical
systems analysis of ref.\ \cite{SM1}: any further constraints
would mean we no longer had a primary hair in contradiction with the
results derived there. We will argue that no further constraints
exist for $N>2$ either. In particular, if further constraints exist
on the charges then it would be possible to extract them from the
field equations. If we consider the field equations at spatial infinity,
then solving order by order in inverse powers of $r$ no constraints
on the charges $\sc{a}$ are found, though we do find that all
coefficients of terms $\OO(r^{-n})$, $n\ge2$, in the asymptotic series
(\ref{asymA})--(\ref{asymC}) are completely determined. At
the next order, for example,
\bea
u_{\scrscr2}&=&\bQ,\lb{asym2A}\\
\ph{a}_{\scrscr2}&=&M\sc{a}-\frac12\gg{a}\bQ_a,\lb{asym2B}\\
\Ra2&=&-\SUM1\sc{a}^2.\lb{asym2C}\eea
Solutions with the asymptotic expansions (\ref{asymA})--(\ref{asymC})
include many which correspond to naked singularities rather than black
holes. The requirement that solutions also have a regular horizon leads
to the further constraints (\ref{screl2}), (\ref{constraint}) found
upon integrating the independent field equations from $r=r_+$ to spatial
infinity, as above. However, we can obtain no more than one constraint
for each independent field equation (\ref{febA}), (\ref{febB}), and the
relations (\ref{screl2}), (\ref{constraint}) which give one constraint
for each equation, exhaust the possibilities. Thus we find that there
are $N-1$ independent parameters among the $N$ scalar charges, $\sc{a}$.

\subsection{Thermodynamic quantities}
Even in the absence of complete analytic solutions, some thermodynamic
relations may be obtained, given that the black hole temperature and entropy
are defined at the horizon, $r_+$, which is related to the mass by
(\ref{rpm1}). In particular, let us evaluate the derivative of
(\ref{RU2}) at the horizon. In terms of the surface gravity,
$\kappa=\frac12 \left.\left(\e^{2U}\right)'\right|_{r=r_+}$,
the horizon area $\AH=4\pi R^2(r_+)$, and using substituting for $r_+$
from (\ref{rpm1}) we then find
\beq M={\ka\AH\over4\pi}+\SUM1{\sc{a}\over\gg{a}}\,.\lb{thermo1}\eeq
We now define a magnetostatic potential, $\chi(r)$, according to
\bea -\pt_r\chi&=&\left[\SUM1(\la{a})^2\eph\right]\,{}^*\!F_{tr}\\
&=&\left[\SUM1(\la{a})^2\eph\right]{Q\over R^2}\,.\lb{magnet}\eea
We integrate (\ref{magnet}) from $r=r_+$ to $r=\infty$, and use
(\ref{screl}) to find that the magnetostatic potential of the horizon is
given by
\beq\chiH\equiv\chi(r_+)={1\over Q}\SUM1{\sc{a}\over\gg{a}}\,.\eeq
It then follows that (\ref{thermo1}) is equivalent to the Smarr-type
relation
\beq M={\ka\AH\over4\pi}+Q\chiH\,.\eeq

For completeness, we also note that according to (\ref{rpm1}), the
mass $M$ is a homogeneous function of degree $\frac12$ in $\AH$, and
of degree one in each $\sc{a}$. The appropriate first law of black
hole mechanics for the system is therefore
\beq \dd M=T\dd S+\SUM1{\dd\Sigma_a\over\gg{a}}\lb{first_law}\eeq
where we have identified the temperature, $T=\kappa/(2\pi)$, and
entropy, $S=\frac14\AH$, in the usual fashion.
There is no independent variation $\dd Q$ in (\ref{first_law}), since
$Q$ is related to the scalar charges via the constraint (\ref{constraint}).
Indeed, for adiabatic variations
\beq \dd Q={1\over\chiH}\SUM1{\dd\Sigma_a\over\gg{a}}
\eeq
Interestingly enough, given the form of eq.~(\ref{rpm1}) for $r_+$,
there is no further contribution to the first law from independent
variations of the vacuum moduli charges, $\phm{a}$, as there is in the
case of other theories such as those of refs.~\cite{smarr}. In view
of the relations (\ref{screl2}) this is to be expected.

\section{Linear stability}
In many cases hairy black holes are unstable. In this
section we examine whether this is the case also for our solutions.
It turns out that our solutions are stable against
linear radial perturbations. The general perturbations of the
solutions in the case of a single scalar field (dilaton) has been
studied in great detail in \cite{HW}, using the methods of ref.
\cite{chan}, and there stability has been proved.
The calculations were, however, already very involved in that relatively
simple case, and hence, following most of the
literature on the subject \cite{stab,VG2},
we prefer to limit ourselves to the study of radial perturbations.
Moreover, we shall consider only the case of two scalar fields.

We consider the action (\ref{action2}),
and for simplicity we put $q=1$, $\lambda_2=1$, since considering
the more general set of coupling parameters in the $N=2$ case of
eq.~(\ref{action}) does not affect our conclusions.

For the discussion of stability, it is convenient to use
coordinates in which the metric takes the form \beq\lb{metric} \dd
s^2=-\e^{\Gamma(R,t)}\dd t^2+\e^{\LA(R,t)}\dd
R^2+R^2\dd\Omega^2_2, \eeq with \beq
\Phi=\Phi(t,R),\qquad\Psi=\Psi(t,R), \eeq and the magnetic field
is given in an orthonormal basis by \beq\lb{em}
F_{\hat\th_1\hat\th_2}=Q\,\ep_{\hat\th_1\hat\th_2} \eeq In these
coordinates, the field equations read \bea
&&\Phi''+\conn\Phi'-\elg\left(\ddot\Phi+\cons\dot\Phi\right)\nn\\
&&\quad=-{Q^2\over R^4}\elf,\\ &&\nn\\
&&\Psi''+\conn\Psi'-\elg\left(\ddot\Psi+\cons\dot\Psi\right)\nn\\
&&\quad=-{Q^2\over R^4}\els,\\ &&\nn\\
&&\LA'=R\left[\Phi'^2+\Psi'^2+\elg(\dot\Phi^2+\dot\Psi^2)\right]+
{1-\e^\LA\over R}\nn\\ &&\quad+{Q^2\over
R^3}\left(\elf+\els\right),\\ &&\nn\\
&&\Gamma'=R\left[\Phi'^2+\Psi'^2+\elg(\dot\Phi^2+\dot\Psi^2)\right]-
{1-\e^\LA\over R}\nn\\ &&\quad-{Q^2\over
R^3}\left(\elf+\els\right),\\ &&\nn\\
&&\dot\LA=2R(\dot\Phi\Phi'+\dot\Psi\Psi'),\\ &&\nn\\
&&\Gamma''+\left({\Gamma'\over2}+{1\over
R}\right)(\Gamma'-\LA')-\elg
\left(\ddot\LA+{\dot\LA-\dot\Gamma\over2}\right)\dot\LA\nn\\
&&\quad=2\left[\elg(\dot\Phi^2+\dot\Psi^2)-
(\Phi'^2+\Psi'^2)\right]\nn\\ &&\quad\quad+{2Q^2\over
R^4}\left(\elf+\els\right), \eea where the prime and the dot
denote differentiation with respect to $R$ and $t$, respectively.

We perturb the field equations by time-dependent linear
perturbations of the form \bea
\Gamma(R,t)&=&\Gamma(R)+\dG(R,t)\e^{i\omega t},\nn\\
\LA(R,t)&=&\LA(R)+\dL(R,t)\e^{i\omega t},\nn\\
\Phi(R,t)&=&\Phi(R)+\dP(R,t)\e^{i\omega t},\nn\\
\Psi(R,t)&=&\Psi(R)+\dS(R,t)\e^{i\omega t},\nn \eea where the
perturbations are assumed small and the functions $\Gamma(R)$,
$\LA(R)$, $\Phi(R)$ and $\Psi(R)$ denote the time-independent
unperturbed solutions of the field equations. We did not perturb
the Maxwell field since the electromagnetic Bianchi identities
imply that the monopole-like solution (\ref{em}) must be
independent of the radial coordinate.

The perturbed equations then read \bea
&&\dP''+\conn\dP'+{\Phi'\over2}(\dG'-\dL')-\elg\delta\ddot\Phi\nn\\
&&\quad=-{Q^2\over R^4}\elf(\dL-2\dP),\\ &&\nn\\
&&\dS''+\conn\dS'+{\Psi'\over2}(\dG'-\dL')-\elg\delta\ddot\Psi\nn\\
&&\quad=-{Q^2\over R^4}\els(\dL-2\dS),\\ &&\nn\\
&&\dL'-2R(\Phi\dP'+\Psi\dS')+{\e^\LA\over R}\dL\nn\\
&&\quad={Q^2\over
R^3}\left[\elf(\dL-2\dP)+\els(\dL-2\dS)\right],\\ &&\nn\\
&&\dG'-2R(\Phi\dP'+\Psi\dS')-{\e^\LA\over R}\dL\nn\\
&&\quad=-{Q^2\over
R^3}\left[\elf(\dL-2\dP)+\els(\dL-2\dS)\right],\nn\\ &&\\
&&\delta\dot\LA=2R(\Phi'\delta\dot\Phi+\Psi'\delta\dot\Psi)\lb{lambda},\\
&&\nn\\ &&\dG''+\left(\Gamma'-{\LA'\over2}+{1\over
R}\right)\dG'-\left({\Gamma'\over2}+{1\over R}
\right)\dL'-\elg\delta\ddot\LA\nn\\
&&\quad=-4(\Phi'\dP'+\Psi'\dS')+{2Q^2\over
R^4}\Big[\elf(\dL-2\dP)\nn\\ &&\quad\quad+\els(\dP-2\dS)\Big].
\eea

Eq. (\ref{lambda}) can be immediately integrated. With suitable
boundary conditions it yields \beq \dL=2R(\Phi'\dP+\Psi'\dS). \eeq

The problem of stability can then be reduced to the study of the
perturbation of the scalar fields $\Phi$ and $\Psi$
\cite{stab,VG2}. After long manipulations of the perturbed
equations, one can obtain a coupled system of second order linear
equations for $\dP$ and $\dS$: \bea
&&\dP''+\conn\dP'+A(R)\dP+C(R)\dS\nn\\
&&\quad=\elg\delta\ddot\Phi,\lb{dp}\\ &&\nn\\
&&\dS''+\conn\dS'+C(R)\dP+B(R)\dS\nn\\
&&\quad=\elg\delta\ddot\Psi,\lb{ds} \eea where \bea
A(R)&=&-2R\left[2\Phi'\Phi''+\cont\Phi'^2\right]\nn\\
&&+2\left[\Phi''+\conn\Phi'\right],\\ &&\nn\\
B(R)&=&-2R\left[2\Psi'\Psi''+\cont\Psi'^2\right]\nn\\
&&+2\left[\Psi''+\conn\Psi'\right],\\ &&\nn\\
C(R)&=&-2R\left[\Phi'\Psi''+\Phi''\Psi'+\cont\Phi'\Psi'\right].\nn\\
\eea

In Schwarzschild coordinates (\ref{metric}) the previous equations
are not regular at the horizon. Therefore, it necessary to define new
``tortoise'' coordinates, given by \cite{VG2}
\bd
R^*=\int\e^{(\Gamma-\LA)/2}\dd R.
\ed
Defining new fields
\bd
u=R\,\dP,\qquad v=R\,\dS,
\ed
and using the explicit time-dependence of the perturbative modes,
one can finally put the stability equations (\ref{dp})-(\ref{ds})
in the Schr\"odinger form
\beq\lb{schr}
{d^2\bu\over \dd R^{*2}}+\omega^2\bu=\bA\bu,
\eeq
where $\bu$ is the vector of components $(u,v)$ and $\bA$ is a
symmetric matrix with entries
\bea
&&A_{11}=-\left(A-{\Gamma'-\LA'\over2R}\right)\egl,\nn\\
&&A_{12}=A_{21}=-C\egl,\nn\\
&&A_{22}=-\left(B-{\Gamma'-\LA'\over2R}\right)\egl.\nn
\eea

The matrix $\bA$ can be diagonalized, and has eigenvalues
\bd
V_{1,2}=\ha\left[(A_{11}+A_{22})\pm\sqrt{(A_{11}-A_{22})^2+4A_{12}^2}\right].\nn
\ed
The classical solutions are stable under linear perturbations if the potentials
$V_{1,2}$ are everywhere positive.
This can be proved by generalizing the arguments of Chandrasekhar
\cite{chan}. In fact, (\ref{schr}) can be written as
\beq\lb{sch}
{\de^2\bu\over\de t^2}-{\de^2\bu\over\de R^{*2}}+\bA\bu=0.
\eeq
Multiplying (\ref{sch}) by $\de\bu^\dagger/\de t$ and integrating over
$R^*$, one gets
\bd
\int\left({\de\bu^\dagger\over\de t}{\de^2\bu\over\de t^2}-
{\de\bu^\dagger\over\de t}{\de^2\bu\over\de R^{*2}}+
{\de\bu^\dagger\over\de t}\bA\bu\right)\dd R^*.
\ed
After integrating the second term by parts, and adding the complex
conjugate equation, one obtains the energy integral
\beq
\int\left(\left|\de\bu\over\de t\right|^2+\left|\de\bu\over\de
R^*\right|^2+\bu^\dagger\bA\bu\right)\dd R^*=\const.
\eeq
If the last two terms in the integral are bounded and positive definite,
it follows that the integral of $|\de\bu/\de t|^2$ is also bounded, ruling
out any exponential growth of the perturbations. In our case, it is
sufficient to show that $\bu^\dagger\bA\bu$ is positive. This can
be easily checked by diagonalizing the matrix $\bA$. If the
eigenvalues $V_{1,2}$ are non-negative functions, then
$\bu^\dagger\bA\bu$ is clearly positive.

\begin{figure}[htb]
\vbox{ \vskip 10 pt
\centerline{\scalebox{0.75}{\includegraphics{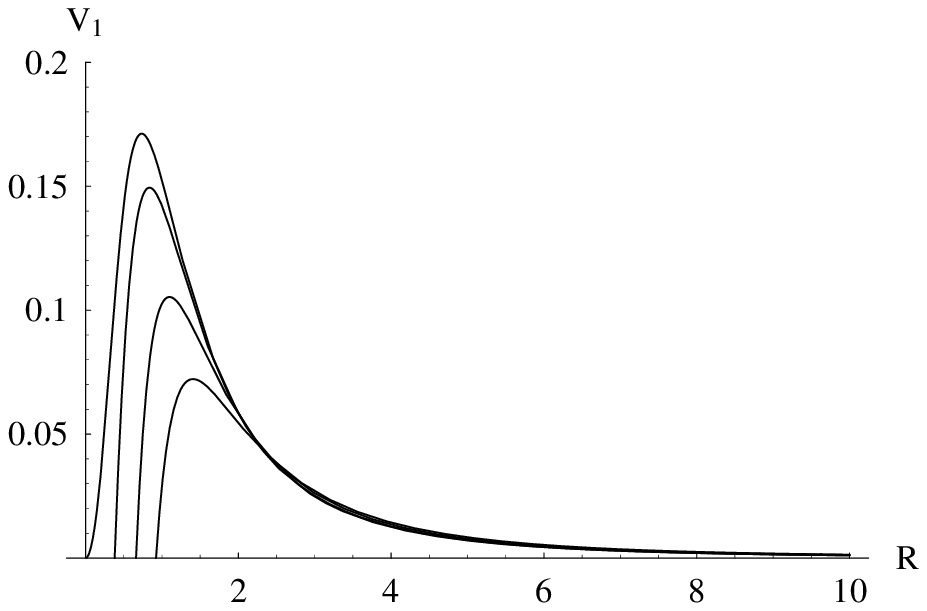}}}
\vskip 10pt
\centerline{\scalebox{0.75}{\includegraphics{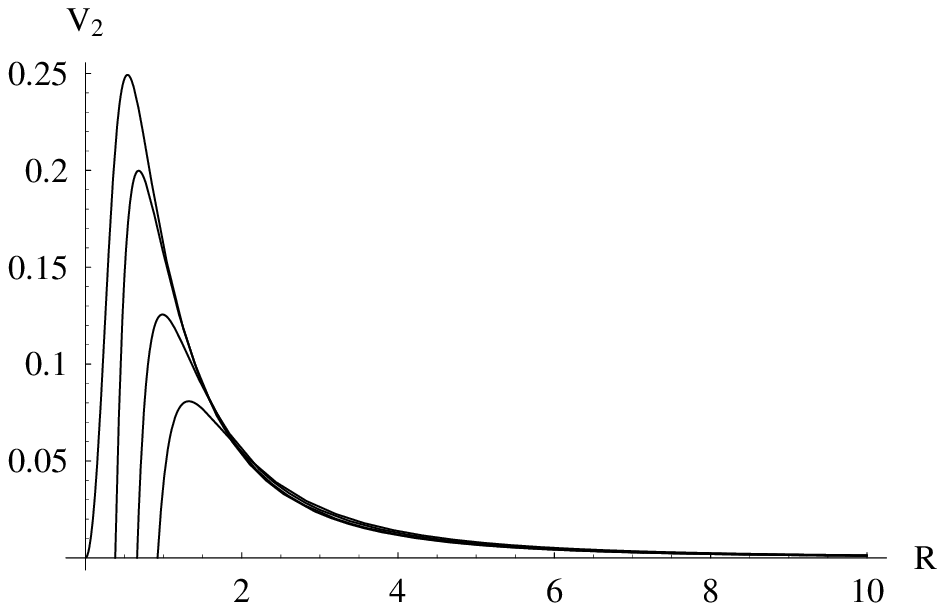}}}
\caption{%
{\sl The potentials $V_1$ and $V_2$ for the exact solution, with
different values of $b/a$, and hence
of the horizon. The extremal case corresponds to horizon at $R=0$.}}}
\end{figure}

This can be readily checked for the exact solutions. In the other
cases, one has of course to resort to numerical calculations. Let
us consider for example the solution with $\Phi=\Psi$ \cite{SM1}.
This can be written as \bea &&\dd s^2= -{(r-\rp)(r-\rq)^{1/3}\over
r^{4/3}}\dd t^2+{r^{4/3}\;\dd r^2 \over(r-\rp)(r-\rq)^{4/3}}\nn\\
&&\quad\qquad+r^{2/3}(r-\rq)^{4/3}\dd\Omega^2,\\ &&\nn\\
&&\ef=\es=\left(1-{\rq\over r}\right)^{2/3}. \eea In terms of the
coordinate $R$ such that the metric takes the form (\ref{metric}),
one has \bea
&&\e^\Gamma={(a^2-2a\De+\De^2)^{1/3}(a-b\De+\De^2)\over
(a^2+\De+\De^2)^{4/3}},\\ &&\nn\\ &&\e^\LA=\nn\\
&&\quad{R(a^2-\De^2)^2(a^2+a\De+\De^2)^{1/3}\over(R^3+4a^3)
\De^2(a^2-2a\De+\De^2)^{1/3}(a^2-b\De+\De^2)},\nn\\ &&\\
&&\ef=\es=\left({a^2-2a\De+\De^2\over
a^2+a\De+\De^2}\right)^{2/3}, \eea where $a=\rq/3$, $b=\rp-\rq/3$,
and \beq \De=\ ^3\sqrt{R^3+2a^3+\sqrt{R^3(R^3+4a^3)}}.\nn \eeq In
these coordinates, the singularity is located at $R=0$, and the
horizon at $R=(b+a)^{2/3}(b-2a)^{1/3}$.

One can now substitute the metric functions in (\ref{schr}). The
analysis is greatly simplified by the fact that for this solution
$\Phi=\Psi$ and hence $A(R)=B(R)$. The equations readily separate
into two independent equations for $u+v$ and $u-v$, with
potentials $V_1=A_{11}+A_{12}$ and $V_2=A_{11}-A_{12}$,
respectively.

The potentials are plotted for different values of the ratio
$\rp/\rq$ in fig.\ 2. They vanish at the horizon and at
infinity and are regular and positive in the interval. We can
hence deduce the stability of the solution with $\Phi=\Psi$.

In the general case, numerical calculations show that the
behaviour of $V_{1,2}$ is qualitatively the same as for exact
solutions. Some examples are given in fig.\ 3. We can conclude
that all the classical solutions are stable against radial linear
perturbations.

\begin{figure}[htb]
\vbox{
\vskip 10 pt
\centerline{\scalebox{0.75}{\includegraphics{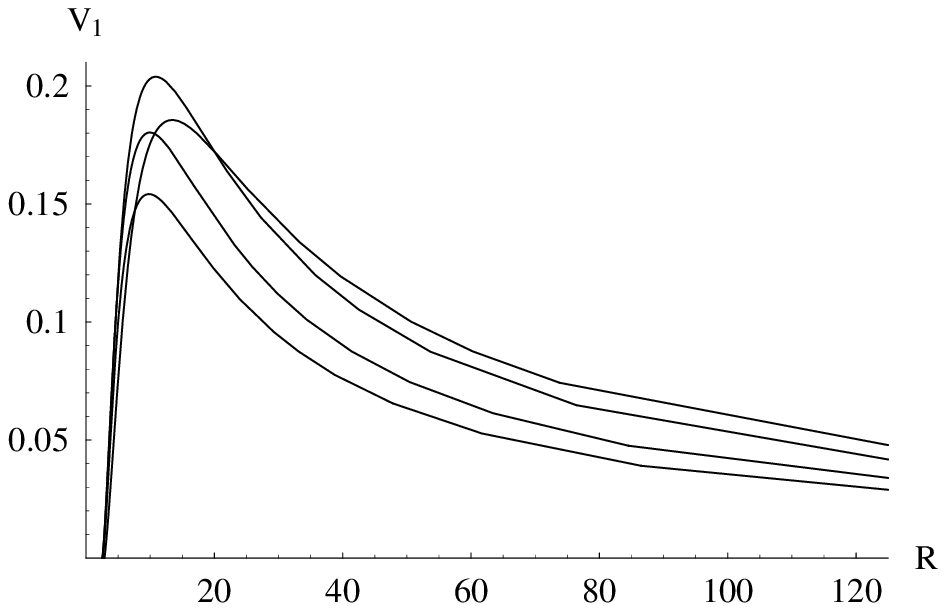}}}
\vskip 10 pt
\centerline{\scalebox{0.75}{\includegraphics{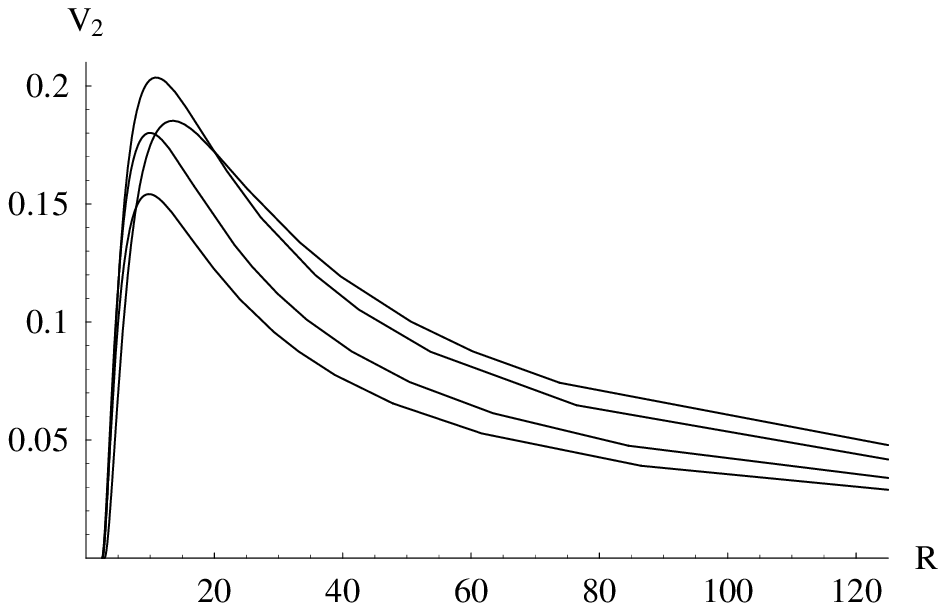}}}
\caption{%
{\sl The potentials $V_1$ and $V_2$ for numerical solutions
with fixed values of $r_+$ and $r_-$, and variable third independent
parameter. The height of the peak increases with the value of
the ratio between the scalar charges. The lowest one corresponds to
$\Si_\Phi=\Si_\Psi$.}}}
\end{figure}

\section{Discussion}

We have generalized the solutions previously found in ref.\ \cite{SM1}
for magnetically charged black holes holes non-minimally coupled to
two scalar fields, to the case of multiple scalar fields non-minimally
coupled to a single magnetic monopole. Even though the complete analytic
solutions have not been derived, we have succeeded
in integrating enough of the field equations that constraints on the
masses and charges can be derived. This analysis supports the claim
made in ref.\ \cite{SM1} that the solutions possess a primary hair.
In the case of $N$ scalar fields there are $N-1$ independent parameters
among the scalar charges.

We have further shown that in the case of two scalar fields, that the
black hole solutions are classically stable to radial perturbations, a
feature which is absent in the case of a number of other hairy black
hole solutions. We have no reason to expect that the case of multiple
scalar fields as given in Sec.~II would be any different.

Our present analysis indicates that the primary hair of the multiscalar
black holes has quite novel features as compared to the case of other hairy
black holes. In particular, while the black hole is characterized by new
independent degrees of freedom which are defined at spatial infinity, these
charges must necessarily vanish if the magnetic field is turned off. To
distinguish such scalar hair from the case of {\it elementary primary
scalar hair}, which would theoretical exist even in the absence of
gauge charges, we identify this new form of hair as {\it contingent
primary scalar hair}. This suggests a further refinement of the no
hair theorems by the statement: {\it In theories satisfying the DEC,
black holes do not possess elementary primary scalar hair}.

Our analysis of the thermodynamic properties of the solutions has been
limited to the derivation of a Smarr formula, and of the first law.
A further step would be the thorough study of the
thermodynamical properties of the solutions using explicit numerical
solutions.

We also remark that solutions with properties analogous to those
of the model studied in this paper have been
obtained in the case of Gauss-Bonnet gravity non-minimally coupled
to two scalar fields \cite{AM}. It would be interesting to investigate
if our results can be extended also to that case.

\begin{acknowledgments}
DLW would like to thank Gary Gibbons and Ian Moss for helpful discussions
and also the Mathematics Department at the University of Cagliari, and the
INFN, Cagliari for their hospitality, when this work was initiated.
SM would like to thank the Department of Physics and Astronomy of the
University of Canterbury for their hospitality during later stages of this
work. This work was financially supported by a coordinate research project
of the University of Cagliari, by Australian Research Council grant
F6960043, by Adelaide University Small Grant 21060100, and by the
Marsden Fund of the Royal Society of New Zealand.
\end{acknowledgments}

\def\PRL#1{Phys.\ Rev.\ Lett.\ {\bf#1}} \def\PR#1{Phys.\ Rev.\ {\bf#1}}
\def\CQG#1{Class.\ Quantum Grav.\ {\bf#1}} \def\JP#1{J.\ Phys.\ {\bf#1}}
\def\AP#1#2{Ann.\ Phys.\ (#1) {\bf#2}} \def\NP#1{Nucl.\ Phys.\ {\bf#1}}
\def\JMP#1{J.\ Math.\ Phys.\ {\bf#1}}
\def\PL#1{Phys.\ Lett.\ {\bf#1}} \def\MPL#1{Mod.\ Phys.\ Lett.\ {\bf#1}}
\def\CMP#1{Commun.\ Math.\ Phys.\ {\bf#1}}
\def\GRG#1{Gen.\ Relativ.\ Grav.\ {\bf#1}}
\def\IJMP#1{Int.\ J. Mod.\ Phys.\ {\bf #1}}


\begin{thebibliography}{66}

\bibitem{Be}
J.D. Bekenstein,
\PR{D5}, 1239 (1972);
\PR{D5}, 2403 (1972).

\bibitem{Be1}
J.D. Bekenstein,
\AP{NY}{82}, 535 (1974);
\AP{NY}{91}, 75 (1975).

\bibitem{SZ}
D. Sudarsky and T. Zannias,
\PR{D58}, 087502 (1998). 

\bibitem{LM}
H. Luckock and I.G. Moss,
\PL{B176}, 341 (1986);
H. Luckock,
in N. Sanchez and H. de Vega (eds.) {\it``String theory, quantum
cosmology and quantum gravity, integrable and conformal invariant
theories''} (World Scientific, Singapore) p.\ 454; S. Droz, M.
Heusler and N. Straumann,
\PL{B268}, 371 (1991);
I.G. Moss, N. Shiiki and E. Winstanley,
\CQG{17}, 4161 (2000). 

\bibitem{EYMH}
M.E. Ortiz,
\PR{D45}, R2586 (1992);
K. Lee, V.P. Nair and E.J. Weinberg,
\PR{D45}, 2751 (1992); P. Breitenlohner, P. Forg\'acs and D.
Maison,
\NP{B383}, 357 (1992);
\NP{B442}, 126 (1995).

\bibitem{VG}
M.S. Volkov and D.V. Gal'tsov,
Phys.\ Rep.\ {\bf319}, 1 (1999). 

\bibitem{BL}
O. Bechmann and O. Lechtenfeld,
\CQG{12}, 1473 (1995); 
H. Dennhardt and O. Lechtenfeld,
\IJMP{A13}, 741 (1998). 

\bibitem{BS}
K.A. Bronnikov and G.N. Shikin,
Gravit.\ Cosmol.\ {\bf8}, 313 (2002). 

\bibitem{NS}
U. Nucamendi and M. Salgado,
\PR{D68}, 044026 (2003). 

\bibitem{DM}
P. Dobiasch and D. Maison,
\GRG{14}, 231 (1982);
G.W. Gibbons and D.L. Wiltshire,
\AP{N.Y.}{167}, 201 (1986); (E) {\bf176}, 393 (1987).

\bibitem{GM}
G.W. Gibbons and K. Maeda,
\NP{B298}, 741 (1988);
D. Garfinkle, G.T. Horowitz and A. Strominger,
\PR{D43}, 3140 (1991); (E) {\bf D45}, 3888 (1992).

\bibitem{MS}
S. Mignemi and N. Stewart,
\PR{D47}, 5259 (1993).

\bibitem{CPW1}
S. Coleman, J. Preskill and F. Wilczek,
\PRL{67}, 1975 (1991);
A. Shapere, S. Trivedi and F. Wilczek,
\MPL{A6}, 2677 (1991).

\bibitem{CPW2}
S. Coleman, J. Preskill and F. Wilczek,
\NP{B378}, 175 (1992).

\bibitem{SM1}
S. Mignemi,
\PR{D62}, 024014 (2000). 

\bibitem{notnote}
In the notation of \cite{SM1} $\si\equiv\Psi/\sqrt{3}$.

\bibitem{MW}
S. Mignemi and D.L. Wiltshire,
\CQG{6}, 987 (1989);
D.L. Wiltshire,
\PR{D44}, 1100 (1991);
S. Mignemi and D.L. Wiltshire,
\PR{D46}, 1475 (1992); 
S.J. Poletti and D.L. Wiltshire,
\PR{D50}, 7260 (1994); (E) {\bf D52}, 3753 (1995). 

\bibitem{BMG}
P. Breitenlohner, D. Maison and G.W. Gibbons,
\CMP{120}, 295 (1988).

\bibitem{CC}
M. Cadoni and C.N. Colacino,
\NP{B590}, 252 (2000).

\bibitem{PTW}
S.J. Poletti, J. Twamley and D.L. Wiltshire,
\PR{D51}, 5720 (1995); 
\CQG{12}, 1753 (1995), (E) {\bf12}, 2355 (1995); 
D.L. Wiltshire,
J.\ Austral.\ Math.\ Soc.\ {\bf B41}, 198 (1999). 

\bibitem{stab}
N.~Straumann and Z.H.~Zhou,
\PL{B237}, 353 (1990);
\PL{B243}, 33 (1990);
M. Heusler, S. Droz and N. Straumann, \PL{B271}, 61 (1991);
\PL{B285}, 21 (1992);
E. Winstanley and N.E. Mavromatos,
\PL{B352}, 242 (1995).

\bibitem{VG2}
M.S. Volkov and D.V. Gal'tsov,
\PL{B341}, 279 (1995).

\bibitem{HW}
C.F.E. Holzhey and F. Wilczek,
\NP{B380}, 447 (1992).

\bibitem{KW}
P. Kanti, N.E. Mavromatos, J. Rizos, K. Tamvakis and E. Winstanley,
\PR{D57}, 6255 (1998).

\bibitem{smarr}
D.A. Rasheed,
{\tt hep-th/9702087};
M. Heusler and N. Straumann,
\CQG{10}, 1299 (1993);
G.W. Gibbons, R. Kallosh and B. Kol,
\PRL{77}, 4992 (1996).

\bibitem{chan}
S. Chandrasekhar, {\it The mathematical theory of black holes},
(Clarendon Press, Oxford, 1983).

\bibitem{AM}
S. Alexeyev and S. Mignemi,
\CQG{18}, 4165 (2001).
\end{thebibliography}
\end{document}